\documentstyle[preprint,aps]{revtex}

\newcommand{{\calA}}{\mbox{\boldmath$\cal A$\unboldmath}}
\newcommand{{\bfchi}}{\mbox{\boldmath$\chi$\unboldmath}}
\newcommand{{\Beta}}{\mbox{\boldmath$\beta$\unboldmath}}
\newcommand{{\bfpi}}{\mbox{\boldmath$\pi$\unboldmath}}
\newcommand{{\bfmu}}{\mbox{\boldmath$\mu$\unboldmath}}
\newcommand{{\bfsigma}}{\mbox{\boldmath$\sigma$\unboldmath}}
\newcommand{{\bftau}}{\mbox{\boldmath$\tau$\unboldmath}}
\newcommand{{\bfa}}{\mbox{\boldmath$a$\unboldmath}}
\newcommand{{\bfE}}{\mbox{\boldmath$E$\unboldmath}}
\newcommand{{\bfB}}{\mbox{\boldmath$B$\unboldmath}}
\newcommand{{\bfF}}{\mbox{\boldmath$F$\unboldmath}}
\newcommand{{\bfv}}{\mbox{\boldmath$v$\unboldmath}}
\newcommand{{\bfV}}{\mbox{\boldmath$V$\unboldmath}}
\newcommand{{\bfA}}{\mbox{\boldmath$A$\unboldmath}}
\newcommand{{\bfO}}{\mbox{\boldmath$O$\unboldmath}}
\newcommand{{\bfJ}}{\mbox{\boldmath$J$\unboldmath}}
\newcommand{{\bfx}}{\mbox{\boldmath$x$\unboldmath}}
\newcommand{{\bfR}}{\mbox{\boldmath$R$\unboldmath}}
\newcommand{{\bfr}}{\mbox{\boldmath$r$\unboldmath}}
\newcommand{{\bff}}{\mbox{\boldmath$f$\unboldmath}}
\newcommand{{\bfd}}{\mbox{\boldmath$d$\unboldmath}}
\newcommand{{\bfh}}{\mbox{\boldmath$h$\unboldmath}}
\newcommand{{\bfp}}{\mbox{\boldmath$p$\unboldmath}}
\newcommand{{\bfk}}{\mbox{\boldmath$k$\unboldmath}}
\def\ddot#1{\raise6.0pt\hbox{{\bf\tenrm {\bf ..}}}\mkern-12.0mu{#1}}
\def\DDot#1{\raise8.0pt\hbox{{\bf\tenrm {\bf ..}}}\mkern-14.0mu{#1}}
\def\DDDot#1{\raise8.0pt\hbox{{\bf\tenrm {\bf ...}}}\mkern-16.0mu{#1}}
\def\dddot#1{\raise6.0pt\hbox{{\bf\tenrm {\bf ...}}}\mkern-14.5mu{#1}}
\def\v#1{{\bf#1}}
\def\'#1{{\accent19\ifx #1i \i\else #1\fi}}

\def \oe{I\!E}
\def \ob{I\!B}

\begin{document}
\draft
\title{\huge\bf\textsf {The Kirchhoff gauge}}
\author{{\bf Jos\'e  A. Heras$^*$}\\ 
{\small \em Departamento de F\'isica, E.S.F.M., Instituto Polit\'ecnico Nacional, M\'exico D. F., M\'exico \\ 
and Department of Physics and Astronomy, Louisiana State University, Baton Rouge, LA 70803-4001, USA}\\
\vspace*{28pt}
\parbox{141mm}{\normalsize {\bf Abstract}
\vskip 4pt We discuss the Kirchhoff gauge in classical electrodynamics. In this gauge the scalar potential satisfies an elliptical equation and the vector potential satisfies a wave equation with a nonlocal source. 
We find the solutions of both equations and show that, despite of the unphysical character of the scalar potential, the electric and magnetic fields obtained from the scalar and vector potentials are given by their well-known retarded expressions. We note that the Kirchhoff gauge pertains to the class of gauges known as the velocity gauge. 
**e-mail: heras@phys.lsu.edu}}
%\vskip 14pt
\maketitle

\newpage
\noindent{\bf 1. Introduction}
\vskip 4pt
The long history that led to the conclusion that the scalar and vector potentials in electromagnetism are not unique 
and that different potentials connected by a gauge transformation describe the same physical fields has recently been reviewed by Jackson and Okun [1]. 
In a further pedagogical paper, Jackson [2] derived some explicit gauge functions that transform potentials in a gauge into potentials in another gauge and emphasized that, whatever propagation or nonpropagation characteristics of potentials in a particular gauge, the electric and magnetic fields are always the same. 
According to the authors of Ref. 1, the first published relation between potentials is due to Kirchhoff [3] who showed that the Weber form of the vector potential $\v A$ and its associated scalar potential $\Phi$ satisfy the relation (in modern notation):
\begin{eqnarray}
\nabla\cdot\v A=\frac{1}{c}\frac{\partial\Phi}{\partial t}.
\end{eqnarray}
Clearly, Eq. (1) is a gauge condition which will be called here the Kirchhoff condition. The associated gauge will be called the Kirchhoff gauge. Apparently, Eq. (1) was obtained only for quasistatic potentials in which retardation is neglected [2]. A general discussion on the Kirchhoff gauge does not seem to have been presented yet in the literature.

In this paper we discuss the Kirchhoff gauge. We show that in this gauge the scalar potential $\Phi_K$ satisfies an elliptical equation, which does not describe a real propagation. This elliptical equation may be rewritten in such a way that it formally states that $\Phi_K$ ``propagates" with an imaginary speed $``ic"$,
 emphasizing even more the unphysical character of $\Phi_K$. Interestingly, we show that $\Phi_K$ has the same structural form that the scalar potential (in the associated Lorenz condition) of an electromagnetic theory formulated in the Euclidean four-space [4].    
We also show that the vector potential in the Kirchhoff gauge $\v A_K$ satisfies a wave equation with a nonlocal source. The solution of this equation involves two cumbersome three-dimensional spatial integrals. Then we proceed to find a more approachable form for $\v A_K$.
By making use of the potential $\Phi_K$ and the Lorenz-gauge scalar potential $\Phi_L$, we find an expression for the gauge function $\chi^{K}$ that connects both potentials. We then use the function $\chi^{K}$ together with the Lorenz-gauge vector potential $\v A_L$ to obtain a transparent expression for the potential $\v A_K$, which involves now one three-dimensional spatial integral and one time integral. We explicitly show that, despite of the unphysical character of $\Phi_K$, the electric and magnetic fields obtained from   $\Phi_K$ and $\v A_K$  are given by their well-known retarded expressions. 
Finally, we point out that the Kirchhoff gauge pertains to the class of gauges known generically as the velocity gauge [2].

\vskip 8pt

\noindent{\bf 2. The Kirchhoff gauge}
\vskip 4pt
It is well-known that the electric and magnetic fields $\v E(\v x,t)$ and $\v B(\v x,t)$ are determined from the scalar and vector potentials $\Phi(\v x,t)$ and $\v A(\v x,t)$: 
 \begin{eqnarray}
\v E=-\nabla\Phi- \frac{1}{c}\frac{\partial\v A}{\partial t},\qquad
\v B=\nabla\times\v A.
\end{eqnarray}
These are invariant under the gauge transformations 
 \begin{eqnarray}
\Phi'=\Phi-\frac{1}{c}\frac{\partial \chi}{\partial t},\quad \v A'= \v A +\nabla\chi,
\end{eqnarray}
where $\chi(\v x,t)$ is an arbitrary gauge function. Here we are using Gaussian units and considering fields with localized sources in vacuum. The inhomogeneous Maxwell equations together with Eqs. (2) lead to the coupled equations:
\begin{eqnarray}
&&\nabla^2\Phi+\frac 1c \frac{\partial}{\partial t}(\nabla\cdot\v A)= -4\pi\rho,\\
&&\nabla^2{\v A}-\frac{1}{c^2}\frac{\partial^2\v A}{\partial t^2} - \nabla\bigg(\nabla\cdot\v A+\frac 1c\frac
{\partial \Phi}{\partial t}\bigg)=-\frac{4\pi}{c}\v J.
\end{eqnarray}
When Eq. (1) is used into Eqs. (4) and (5) we obtain
\begin{eqnarray}
\nabla^2\Phi_K+\frac {1}{c^2} \frac{\partial^2\Phi_K}{\partial t^2}=&& -4\pi\rho,\\
\nabla^2{\v A}_K-\frac {1}{c^2} \frac{\partial^2\v A_K}{\partial t^2}=&&-\frac{4\pi}{c}\bigg(\v J
-\frac{1}{2\pi}\nabla\frac{\partial\Phi_K}{\partial t}\bigg),
\end{eqnarray}
where we have written $\Phi_K$ and $\v A_K$ to denote the Kirchhoff-gauge potentials, {\it i.e.,} the potentials that satisfy the Kirchhoff condition:
\begin{eqnarray}
\nabla\cdot\v A_K-\frac{1}{c}\frac{\partial\Phi_K}{\partial t}=0.
\end{eqnarray}
Equation (6) for the potential $\Phi_K$ is an elliptical equation, which does not describe a real propagation. When $\rho=0$ the solutions of Eq. (6) are either exponentially growing or decaying functions [4]. On the other hand, Eq. (7) is a 
wave equation with a nonlocal source (originated by the second term on its right-hand side).
  
To demonstrate that Eqs. (6)-(8) are equivalent to the Maxwell equations we need to prove that the potentials $\Phi_K$ and $\v A_K$ lead to the well-known retarded solutions of Maxwell's equations. Before making this, let us first prove that we can always find potentials satisfying the Kirchhoff condition.
Suppose that our original potentials satisfy Eqs. (4) and (5) but they do not satisfy Eq. (1), {\it i.e.}, $\nabla\cdot\v A-(1/c)\,\partial \Phi/\partial t\not=0$. Let $\nabla\cdot\v A-(1/c)\,\partial \Phi/\partial t=g,$ where $g=g(\v x, t)$ is some known scalar function. 
Let us make a gauge transformation to potentials $\Phi'$ and $\v A'$ and demand that they satisfy the Kirchhoff condition:
\begin{eqnarray}
\nabla\cdot\v A'-\frac{1}{c}\frac{\partial\Phi'}{\partial t}=0=\nabla^2\chi+\frac {1}{c^2} \frac{\partial^2\chi}{\partial t^2}+ g.
\end{eqnarray}
Therefore, provided a gauge function can be found to satisfy 
\begin{eqnarray}
\nabla^2\chi+\frac {1}{c^2} \frac{\partial^2\chi}{\partial t^2}= -g,
\end{eqnarray}
the new potentials $\Phi'$ and $\v A'$ will satisfy the Kirchhoff condition (1) as well as Eqs. (6) and (7).
Then we need to solve Eq. (10) which is the same form that Eq. (6). 
We note that both elliptical equations can be written, after the simple substitution $c^2\to -[ic]^2$, as equations of the general form 
\begin{eqnarray}
\nabla^2\Psi(\v x,t)-\frac {1}{[ic]^2} \frac{\partial^2\Psi(\v x,t)}{\partial t^2}= -4\pi f(\v x,t),
\end{eqnarray}
which states that $\Psi$ ``propagates" with an imaginary speed $``ic"$. Of course, when the change 
\begin{eqnarray}
c \rightarrow ic,
\end{eqnarray}
is made into a standard wave equation then we obtain an elliptic equation of the form (11).
We can show directly that the infinity-space retarded-imaginary Green function [4]:
\begin{eqnarray}
G_K(\v x,t;\v x',t')=\frac{\delta\{t'-t+R/[ic]\}}{R},
\end{eqnarray}
where $R=|\v x- \v x'|,$ satisfies the elliptical equation 
\begin{eqnarray}
 \bigg(\nabla^2-\frac {1}{[ic]^2} \frac{\partial^2}{\partial t^2}\bigg)G_K(\v x,t;\v x',t')=-4\pi\delta(\v x-\v x')\delta(t-t').
  \end{eqnarray}
With the aid of the function $G_K$, we can solve Eq. (11):
\begin{eqnarray}
\Psi(\v x,t)= \int\int d^3x' dt'G_K(\v x,t;\v x',t')f(\v x',t'),
\end{eqnarray}
where the space integral is over all space and the time integral is from $-\infty$ to $+ \infty$. Time integration
of Eq. (15) gives the more transparent expression:
\begin{eqnarray}
\Psi(\v x,t)= \int d^3x' \frac{f(\v x',t-R/[ic])}{R},
\end{eqnarray}
according to which $\Psi$ propagates with an imaginary velocity $ic$. Of course, if we make the change (12) into the familiar retarded solution of the wave equation then we obtain the retarded-imaginary solution (16).  Therefore, the
existence of the function $\chi$ satisfying Eq. (10) [which has the form of Eq. (11)] is generally guaranteed and thus we can always find potentials satisfying the Kirchhoff condition.

The solution of Eq. (6) can be written as
\begin{eqnarray}
\Phi_K(\v x,t)= \int d^3x' \frac{\rho(\v x',t-R/[ic])}{R}.
\end{eqnarray}
On the other hand, the solution of Eq. (7) is given by
\begin{eqnarray}
{\v A}_K= \frac{1}{c}\int\int G_R \bigg(\v J
-\frac{1}{2\pi}\nabla'\frac{\partial\Phi_K}{\partial t'}\bigg)\;d^3x' dt',
\end{eqnarray}
where 
\begin{eqnarray}
G_R(\v x,t;\v x',t')=\frac{\delta\{t'-t+R/c\}}{R},
\end{eqnarray}
is the well-known infinite-space  Green function
satisfying 
\begin{eqnarray}
 \bigg(\nabla^2-\frac {1}{c^2} \frac{\partial^2}{\partial t^2}\bigg)G_R(\v x,t;\v x',t')=-4\pi\delta(\v x-\v x')\delta(t-t').
  \end{eqnarray}
When we substitute Eq. (17) into Eq. (18) we see that the latter involves two cumbersome three-dimensional spatial integrals. In Sec. 4 we will find a more transparent form for the potential ${\v A}_K$.

\vskip 8pt

\noindent{\bf 3. An alternative interpretation for $\Phi_K$}
\vskip 4pt
An alternative interpretation for the potential $\Phi_K$ may be obtained from an electromagnetic theory formulated in an Euclidean four-space [4]. As is well-known, the  Maxwell equations essentially fix the spacetime signature to be Lorentzian. Electromagnetic fields in a spacetime of Euclidean signature pertain to a theory different from that of Maxwell. The Euclidean fields are seen to satisfy an Euclidean version of Maxwell's equations which was discussed some years ago by Zampino [4] and Brill [5] and more recently by the author [6], kobe [7] and Itin and Hehl [8]. In vector notation the Euclidean equations can be written as
\begin{eqnarray}
\nabla\cdot\rm\oe =&& 4\pi\rho,\\
\nabla\cdot\rm\ob=&&0, \\
\nabla\times\rm\oe-\frac{1}{c}\frac{\partial\rm\ob}{\partial t}=&&0,\\ 
\nabla\times\rm\ob-\frac{1}{c}\frac{\partial\rm\oe}{\partial t}=&&\frac{4\pi}{c}\v J,
\end{eqnarray}
where $\rm\oe$ and $\rm\ob$ are the Euclidean electric and magnetic fields produced by 
the usual charge and current densities, $\rho$ and $\v J$. We are using Gaussian units and considering fields with localized sources in vacuum. Equations (21)-(24) satisfy the continuity equation. By assuming a suitable transformation for the fields $\rm\oe$ and $\rm\ob$, Eqs. (21)-(24) are shown to be invariant under coordinate transformations in a spacetime of Euclidean signature [4]. 
As may be seen, the only formal difference between Maxwell's equations  and the Euclidean equations is the sign of the time derivative of the magnetic field in Eq. (23), or equivalently, the sign in Lenz's law, which appears reversed with respect to the sign of the current Faraday induction law: $\nabla\times\v E+(1/c)\partial\v B/\partial t=0$. Euclidean electrodynamics is just like ordinary electrodynamics except for an ``anti-Lenz" law [5]. 
The change of sign has, however, important consequences. It transforms the associated hyperbolical wave equations to ones of elliptic nature, so that there is no real propagation for Euclidean fields. 

We can also introduce Euclidean potentials to express the Euclidean electric and magnetic fields. 
Equations (22) and (23) allow to write 
\begin{eqnarray}
{\rm \oe}=-\nabla\Phi^E + \frac{1}{c}\frac{\partial \v A^E}{\partial t}, \qquad
{\rm \ob}=\nabla\times\v A^E, 
\end{eqnarray}
where $\Phi^E(\v x,t) $ and $\v A^E(\v x,t)$ are the Euclidean scalar and vector potentials (note the plus sign of the second term in the right-hand side of 
the first relation). 
The fields ${\rm \oe}$ and ${\rm \ob}$ are invariant under the gauge transformations: 
\begin{eqnarray}
\Phi'^E=\Phi^E+\frac{1}{c}\frac{\partial \chi}{\partial t}\qquad \v A'^E= \v A^E+\nabla\chi, 
\end{eqnarray}
where $\chi(\v x,t)$ is an arbitrary gauge function (note the plus sign of the second term in the right-hand side of 
the first relation).  
Equations (21),(24) and (25) imply the coupled partial differential equations:
\begin{eqnarray}
&&\nabla^2\Phi^E-\frac 1c \frac{\partial}{\partial t}(\nabla\cdot\v A^E)= -4\pi\rho,\\
&&\nabla^2{\v A^E}+\frac{1}{c^2}\frac{\partial^2\v A^E}{\partial t^2} - \nabla\bigg(\nabla\cdot\v A^E+\frac 1c\frac
{\partial \Phi^E}{\partial t}\bigg)=-\frac{4\pi}{c}\v J.
\end{eqnarray}
Gauge invariance allows to impose the Lorenz condition on the Euclidean potentials
\begin{eqnarray}
\nabla\cdot \v A_L^E +\frac 1c \frac{\partial\Phi_L^E}{\partial t}=0.
\end{eqnarray}
When Eq. (29) 
is used into Eqs. (27) and (28) we obtain the elliptical equations
\begin{eqnarray}
&&\nabla^2\Phi^E_L+\frac 1{c^2} \frac{\partial^2\Phi^E_L}{\partial t^2}= -4\pi\rho,\\
&&\nabla^2\v A^E_L+\frac{1}{c^2}\frac{\partial^2\v A^E_L}{\partial t^2}=-\frac{4\pi}{c}\v J.
\end{eqnarray}
The solution of Eq. (30) can be written as
\begin{eqnarray}
\Phi_L^E(\v x,t)= \int d^3x' \frac{\rho(\v x',t-R/[ic])}{R}.
\end{eqnarray}
The exact similarity between the Kirchhoff-gauge scalar potential $\Phi_K$ and 
the Lorenz-gauge Euclidean scalar potential $\Phi_L^E$ displayed in Eqs. (6) and (30) [as well as in Eqs. (19) and (32)], allows the identification  $\Phi_K=\Phi_L^E$. Therefore, the Kirchhoff-gauge scalar potential can also be interpreted as 
the Lorenz-gauge potential of an electromagnetic theory formulated in the Euclidean four-space. This emphasizes even more its unphysical character. 

We have here a similar situation to that of the Coulomb gauge in which the scalar potential and its gradient propagate instantaneously which emphasizes its unphysical nature. As is well-known, the instantaneous 
contribution disappears from the final expression of the retarded electric field. This well-known result has recently been emphasized [2, 9, 10, 12]. Similarly, we expect that the unphysical $\Phi_K$ and hence its unphysical gradient $-\nabla\Phi_K$ appearing in the electric field $\v E=-\nabla\Phi_K- (1/c)\partial\v A_K/\partial t$ must exactly cancel with a term arising from the term 
$- (1/c)\partial\v A_K/\partial t$ so that the final result gives the well-known expression of the 
retarded electric field. To show this, it is convenient to find first a more convenient form of $\v A_K $ than that appearing in Eq. (18).
\vskip 8pt

\noindent{\bf 4. An alternative expression for $\v A_K $}
\vskip 4pt
In order to derive an alternative form of Eq. (18), we consider the gauge transformations that convert the Lorenz-gauge potentials, ${\Phi}_L$ and ${\v A}_L$, into the Kirchhoff-gauge potentials, ${\Phi}_K$ and ${\v A}_K$:
\begin{eqnarray}
{\Phi}_K=&& {\Phi}_L -\frac{1}{c}\frac {\partial \chi^K}{\partial t},\\
{\v A}_K=&& {\v A}_L+\nabla \chi^K,
\end{eqnarray}
where $\chi^K$ is the associated gauge function and the potentials $\Phi_L$ and $\v A_L$ are given by 
\begin{eqnarray}
\Phi_L(\v x,t)=&&\int d^3 x'\frac{\rho(\v x',t-R/c)}{R},\\
\v A_L(\v x,t)=&&\int d^3 x'\frac{\v J(\v x',t-R/c)}{Rc}.
\end{eqnarray}
The approach to find $\v A_K$ is similar to one used by Jackson [2] to find an alternative expression for the Coulomb-gauge vector potential. It is relatively simple: using Eqs. (17), (33) and (35) we obtain $\chi^K$. We then calculate its gradient $\nabla \chi^e$ and add it to the potential ${\v A}_L$ to find the potential ${\v A}_K$ using Eq. (34). In fact, from Eqs. (17), (33) and (35) we obtain 
\begin{eqnarray}
\frac{1}{c}\frac{\partial\chi^{K}(\v x,t)}{\partial t}=&& 
\frac 1c\int d^3 x'\frac 1R\bigg[\rho(\v x',t'=t-R/c)
-\rho(\v x',t'=t-R/[ic])\bigg].
\end{eqnarray}
We integrate both sides with respect to {\it ct} to obtain  
\begin{eqnarray}
\chi^{K}(\v x, t)=&& \int d^3 x'\frac cR \bigg[\int_{t_{\bf 0}}^{t-R/c}dt'\;\rho(\v x',t')
-\int_{t_{\bf 0}}^{t-R/({\it i}c)}dt'\;\rho(\v x',t')\bigg] +\chi_{0}.
\end{eqnarray}
This equation can compactly be written as 
\begin{eqnarray}
\chi^{K}(\v x, t)= \int d^3 x'\frac cR \int_{t-R/({\it i}c)}^{t-R/c}dt'\;\rho(\v x',t') +\chi^K_{0}.
\end{eqnarray}
We change variables by writing $t'=t-\tau$ to obtain
\begin{eqnarray}
\chi^{K}(\v x, t)= -\int d^3 x'\frac cR \int_{R/({\it i}c)}^{R/c}d\tau\;\rho(\v x',t-\tau) +\chi^K_{0}.
\end{eqnarray}
The term $\chi^K_{0}$ is constant if we demand finiteness at infinity [2]. The function $\chi^{K}$ in Eq. (40) transforms the Lorenz-gauge potentials into the Kirchhoff-gauge potentials. 
We proceed now to calculate the gradient of Eq. (40). After some direct calculation, we find 
\begin{eqnarray}
\nabla\chi^{K}(\v x,t)=&&\int d^3 x'\bigg[-\frac{\hat{\v R}}{R}\rho(\v x',t')+ \frac{\hat{\v R}}{iR}\rho(\v x',t')\bigg]_{\rm ret({\it c})} +c \int d^3 x'\frac{\hat{\v R}}{R^2}\int_{R/({\it i}c)}^{R/c}d\tau\rho(\v x',t-\tau),
\end{eqnarray}
where ret({\it c}) means $t'= t-R/c$ and $\hat{\v R}=\v R/R$ with $\v R=\v x-\v x'$. Using Eqs. (34), (36) and (41) we find the alternative expression for the Kirchhoff-gauge vector potential:
\begin{eqnarray}
\v A_K(\v x,t)=&&\frac 1c\int d^3 x'\frac{1}{R}\bigg\{\bigg[\v J(\v x',t')-c\hat{\v R}\rho(\v x',t')\bigg]_{\rm ret({\it c})}
+\bigg[\frac{c}{i}\hat{\v R}\rho(\v x',t')\bigg]_{\rm ret({\it ic})} \nonumber\\&&
+\frac{c^2\hat{\v R}}{R} \int_{R/({\it i}c)}^{R/c}d\tau\;\rho(\v x',t-\tau)\Bigg\},
\end{eqnarray}
where ret({\it ic}) means $t'= t-R/[ic].$ Thus, Eq. (18) involving two three-dimensional spatial integrals can alternatively be expressed by Eq. (42) which involves one three-dimensional spatial integral and one time integral replacing the spatial nonlocality of the source with a temporal nonlocality. Of course, Eq. (42) is easier of manipulating than Eq. (18).
The first term in Eq. (42) describes an expected retarded contribution. The second term is an imaginary contribution and the third term mix imaginary and retarded contributions. 

\vskip 8pt

\noindent{\bf 5. The retarded fields}
\vskip 4pt

To verify that the Kirchhoff-gauge potentials lead to the retarded fields we use directly the expressions of the fields in terms of the Kirchhoff-gauge potentials:
 \begin{eqnarray}
\v E=&&-\nabla\Phi_K- \frac{1}{c}\frac{\partial\v A_K}{\partial t},\\
\v B=&&\nabla\times\v A_K.
\end{eqnarray}
We first calculate $-\nabla\Phi_K$ from Eq. (17):
\begin{eqnarray}
-\nabla \Phi_K(\v x,t)&&=\int d^3 x'\frac{1}{R}\Bigg[\frac{\hat{\v R}}{R}\rho(\v x',t')+\frac{\hat{\v R}}{ic}\frac{\partial \rho(\v x',t')}
{\partial t'}\Bigg]_{\rm ret({\it ic})}.%\nonumber\\&&
\end{eqnarray}
Then we calculate $-(1/c)\partial\v A_K/\partial t$ from Eq. (42): 
\begin{eqnarray}
-\frac{1}{c}\frac{\partial\v A_K(\v x,t)}{\partial t}=&&\int d^3 x'\frac{1}{R}\Bigg\{\Bigg[-\frac{1}{c^2}\frac{\partial \v J(\v x',t')}{\partial t'}%\nonumber\\&&
+\frac{\hat{\v R}}{c}\frac{\partial \rho(\v x',t')}{\partial t'}\Bigg]_{\rm ret({\it c})}
-\frac{\hat{\v R}}{ic}\Bigg[\frac{\partial \rho(\v x',t')}{\partial t'}\Bigg]_{\rm ret({\it ic})}\nonumber\\&&
-\frac{\hat{\v R}}{R}\int_{R/({\it i}c)}^{R/c}d\tau
\frac{\partial \rho(\v x',t-\tau)}{\partial t}\Bigg\}.
\end{eqnarray}
With $\partial\rho/\partial t=-\partial\rho/\partial \tau$ we can integrate the last term and so 
Eq. (46) can be written as 
\begin{eqnarray}
-\frac{1}{c}\frac{\partial\v A_K(\v x,t)}{\partial t}=&&-
\int d^3 x'\frac{1}{R}\Bigg[\frac{\hat{\v R}}{R}\rho(\v x',t')%\nonumber\\&&
+\frac{\hat{\v R}}{ic}\frac{\partial \rho(\v x',t')}{\partial t'}\Bigg]_{\rm ret({\it ic})}\nonumber\\&&
+\int d^3 x'\Bigg[\frac{\hat{\v R}}{R^2}\rho(\v x',t')+ \frac{\hat{\v R}}{Rc}\frac{\partial \rho(\v x',t')}{\partial t'}
%\nonumber\\&&
-\frac{1}{Rc^2}\frac{\partial \v J(\v x',t')}{\partial t'}\Bigg]_{\rm ret({\it c})}.
\end{eqnarray}
The first term of Eq. (47) cancels with the term (45) so that 
when Eqs. (45) and (47) are used into Eq. (43) we obtain 
the retarded electric field in the form given by Jefimenko [11]:
\begin{eqnarray}
\v E(\v x,t)=&&\int d^3 x'\Bigg[\frac{\hat{\v R}}{R^2}\rho(\v x',t')+ \frac{\hat{\v R}}{Rc}\frac{\partial \rho(\v x',t')}{\partial t'}%\nonumber\\&&
-\frac{1}{Rc^2}\frac{\partial \v J(\v x',t')}{\partial t'}\Bigg]_{\rm ret({\it c})}.
\end{eqnarray}
On the other hand, Eqs. (42) and (43) give directly the usual expression for the retarded magnetic field: 
\begin{eqnarray}
\v B(\v x,t)=\frac{1}{c}\int d^3x'\frac{1}{R} [\nabla'\times\v J(\v x',t')]_{\rm ret({\it c})}.
\end{eqnarray}
Thus, the Kirchhoff-gauge potentials lead to the familiar retarded fields.

 In Ref. 2 Jackson discussed a class of gauges [12] that he called generically the velocity gauge ($\nu$-gauge) in which the scalar and vector potentials, $\Phi_{\nu}$ and 
$\v A_{\nu}$, are seen to satisfy the gauge condition: 
\begin{eqnarray}
\nabla\cdot\v A_{\nu}+\alpha\frac{1}{c}\frac{\partial\Phi_{\nu}}{\partial t}=0,
\end{eqnarray}
where $\alpha=c^2/\nu^2$. In this gauge the scalar potential propagates with the arbitrary speed $\nu$. The Lorenz and Coulomb gauges are limiting cases, $\alpha=1$ ({\it i.e.,} $\nu=c $) and $\alpha=0$ ({\it i.e.,} $\nu=\infty$), respectively. Jackson [2] emphasized, however, that his: ``...development of the $\nu$-gauge potentials is based on an arbitrary speed of propagation for the scalar potential, either slower or faster than the speed of light, but did not envision an {\it imaginary speed.} That is what happens for a negative $\alpha=c^2/v^2.$" 
We note first that the substitution $\alpha=-1$, which corresponds to the case of the imaginary speed $\nu=ic$, into Eq. (50) [Eq. 7.1 of Ref. 2] gives the Kirchhoff condition (1). What is more important, if we make  
the substitution $\nu=ic$ into Jackson's expression for the $\nu$-gauge vector potential [Eq. 7.8 of Ref. 2] then we  directly obtain Eq. (42). This means that Jackson's treatment of the velocity gauge also covers the case of the imaginary velocity $\nu=ic$. In other words: the Kirchhoff gauge pertains to the class of gauges known generically as the velocity gauge.

\vskip 8pt
\noindent \noindent{\bf Acknowledgment}
\vskip 1pt
\noindent 
%I would like to express my gratitude to Professor R. F. O'Connell for the hospitality extended at   
%the Department of Physics and Astronomy of the Louisiana State University.

The author is grateful to Professor R. F. O'Connell for the kind hospitality extended to him in    
the Department of Physics and Astronomy of the Louisiana State University.

 %to the people of Department of Physics and Astronomy of the Louisiana State University, specially to 
%Professor Robert O'Connell, for their hospitality.


\vskip 5pt

\begin{thebibliography}{99}

\bibitem{1}  
J.D. Jackson, L B Okun, Rev. Mod. Phys. 73 (2001) 663, Available from: $<$hep-ph/0012061$>.$

\bibitem{2} 
J.D. Jackson, Am. J. Phys.  70 (2002) 917, Available from: $<$physics/0204034$>.$ 

\bibitem{3}
G. Kirchhoff, Ann. Phys. Chem. 102 (1857) 529.

\bibitem{4} 
E. Zampino, J. Math. Phys. 27 (1986) 1315.

\bibitem{5} 
D. Brill, in: F. Mansouri, J.J. Escanio (Eds.), Topics on Quantum Gravity and Beyond, World Scientific,
Singapore, 1993, p. 221, Available from: $<$gr-qc/9209009$>.$ 


\bibitem{6} 
J.A. Heras, Am. J. Phys. 62 (1994) 914.

\bibitem{7} 
D.H. Kobe, Am. J. Phys. 65 (1994)  569;\\
D.H. Kobe in: J.A. Heras, R.V. Jim\'enez (Eds.), Topics in Contemporary Physics, IPN, M\'exico (2000), p 63.

\bibitem{8} 
Y. Itin, W. Hehl, Ann. Phys. 312 (2004) 60, Available from: $<$gr-qc/0401016$>.$ 

\bibitem{9}
F. Rohrlich, Am. J. Phys.  70 (2002) 411.

\bibitem{10}
J.A. Heras, Am. J. Phys. 71 (2003) 729;\\
J.A. Heras, Europhys. Lett. 69 (2005)  1.

\bibitem{11}
J.D. Jackson, Classical Electrodynamics, 3rd ed., Wiley, New York, 1999  p. 246.

\bibitem{12}
K.-H. Yang, Ann. Phys. (N.Y.) 168 (1976) 104;\\
K.-H. Yang, Am. J. Phys. 73 (2005) 742.

\end{thebibliography}
\end{document}